\documentclass[preprint]{aastex62}

\graphicspath{{./}{figures/}}

\received{April 27th, 2018}
\revised{November 5th, 2018}
\accepted{November 17th, 2018}
\submitjournal{AJ}

\shorttitle{Ring-Satellite Evolution Regimes and Miranda}
\shortauthors{Hesselbrock and Minton}

\begin{document}

\title{Three Dynamical Evolution Regimes for Coupled Ring-Satellite Systems and Implications for the Formation of the Uranian Satellite Miranda}

\correspondingauthor{Andrew J. Hesselbrock}
\email{ahesselb@purdue.edu, daminton@purdue.edu}

\author[0000-0002-6496-4314]{Andrew J. Hesselbrock}
\affil{Purdue University \\
Department of Physics and Astronomy \\
525 Northwestern Ave. \\
West Lafayette, IN 47907, USA}

\author[0000-0003-1656-9704]{David A. Minton}
\affiliation{Purdue University \\
Department of Earth, Atmospheric, and Planetary Sciences \\
550 Stadium Mall Dr. \\
West Lafayette, IN 47907, USA}



\begin{abstract}
In coupled ring-satellite systems, satellites exchange angular momentum with both the primary through tides and with the ring through Lindblad torques, and may exchange material with the ring through accretion and tidal disruption.
Here we show that these coupled ring-satellite systems fall into three distinct dynamical regimes, which we refer to as ``Boomerang,'' ``Slingshot,'' and ``Torque-Dependent.''
These regimes are determined by the relative locations of the Fluid Roche Limit, the synchronous orbit, and the location of the maximum orbit that Lindblad torques can perturb a satellite.
Satellites that accrete from rings in the Boomerang regime remain interior to the synchronous orbit, and may be driven back toward the primary by tides. 
Satellites that accrete from rings in the Slingshot regime form exterior to the synchronous orbit, and are always driven away from the primary.
Satellites that accrete from rings in the Torque-Dependent regime may exhibit either Boomerang or Slingshot behavior, depending on ring and satellite masses.
We consider both known and hypothesized ring/satellite systems in the solar system, and identify which of these three regimes they fall into.
We determine that Uranus exists within the Torque-Dependent regime.
Using the RING-MOONS code, which models the dynamical evolution of coupled ring-satellite systems, we show that the Uranian satellite Miranda may have accreted from a massive ancient Roche-interior ring and followed a Slingshot-like dynamical path to its present orbit beyond the synchronous orbit, while satellites that accreted after Miranda followed Boomerang-like evolution paths and remained interior to the synchronous orbit.

\end{abstract}



\section{Introduction} \label{sec:intro}

Planetary rings are disks of solid particles in orbit around a primary body, and are common throughout our solar system today.
While Saturn was the first object in the solar system known to have rings \citep{Huygens}, later observations of Jupiter \citep{Smith79}, Uranus \citep{Elliot77}, and Neptune \citep{Hubbard86} revealed that all of the giant planets are orbited by planetary rings.
The giant planets were long thought to be the only objects in the solar system to have rings, however recent discoveries have shown that rings orbit small bodies as well.
Stellar occultation observations suggest that there may be rings in orbit around the centaurs ``10199 Chariklo'' and ``2060 Chiron'' \citep{Braga2014,Ortiz2015}, as well as the dwarf planet Haumea \citep{Ortiz2017}.
These discoveries indicate that rings may be a common feature of many kinds of bodies in the solar system.

Satellite surface processes, giant impacts, tidal disruption, and impact-generated dust and fragments are a few of the processes that may create rings around solar system bodies.
While rings take many forms, we will distinguish between ``tenuous'' ringlets or ``dusty'' belts, and ``massive'' rings \citep{Charnoz2018}.
Tenuous ringlets and dusty belts are strongly influenced by non-gravitational forces.
For instance, ice crystals ejected from active geysers on Saturn's ice satellite Enceladus have collected to form Saturn's E ring \citep{Hamilton94}.
Heliocentric impacts may catastrophically disrupt satellites \citep{Colwell92}, or create dust and fragments that subsequently form these types of rings \citep{Miner2007}.

Massive rings are optically thick and their dynamics are dominated by gravitational, collisional, and tidal processes.
Saturn's main rings are examples of massive rings.
Massive rings can potentially be generated by a variety of processes. 
Close encounters between scattered Kuiper belt objects with the giant planets may have tidally disrupted passing objects into forming rings \citep{Hyodo2017}.
Material ejected from a primary's surface during a giant impact may be placed into orbit, forming a ring \citep{Cameron:1976vi}.
Satellites orbiting close to their primary may also be tidally disrupted to form a ring of material \citep{Canup2010,Black2015,Leinhardt2012Disrupt,Hesselbrock2017}.
Massive rings are the focus of this work.

In addition to the rings known to exist today, massive rings may have once orbited both the Earth and Mars.
A giant impact has been implicated for the formation of Earth's satellite \citep{Cameron:1976vi}.
This impact may have ejected a large amount of material into orbit around Earth, forming a massive ring around the planet, out of which the Moon accreted.

Recent work has hypothesized that Mars may have had a ring system in the past as well.
Large impacts, such as that proposed to have formed the Borealis Basin, could have ejected material into orbit to form a ring \citep{Marinova:2008ia,Citron:2015gb}.
As this ring evolved over time, it may have produced the satellites Phobos and Deimos \citep{Rosenblatt2012,Rosenblatt2016,Hyodo2017,Hesselbrock2017,Canup2018}.
\citet{Hesselbrock2017} (hereafter referred to as HM17) showed that impact debris orbiting Mars may alternate between orbiting Mars as rings or as satellites in an ongoing cycle that eventually produced Phobos.
Thus, HM17 hypothesized that Mars may have had rings at various points in its history.

Planetary rings are composed of solid particles, yet collectively the ring particles behave similar to a fluid.
As the ring particles orbit the primary they continuously collide with other nearby particles, exchanging energy and momentum.
Over time, the intra-ring collisions causes the ring material to ``flow'' from one location to another through a process known as ``viscous spreading'' \citep{Lynden-Bell1974,Goldreich78}.
Intra-ring collisions may also cause some particles to clump together, however for orbits close to the primary, tidal fores disrupt these clumps before they can accrete into satellites.

Ring particles far enough from the primary are able to accrete into satellites \citep{Chandra69}.
Once accreted, these ``ring satellites'' exchange angular momentum with the ring particles.
Perturbations between the ring particles and ring satellites are particularly strong for locations within the ring that are in resonance with the orbiting satellite \citep{Goldreich:1979iq}.
Resonance interactions between ring particles and ring satellites create density waves in the ring which in turn give rise to torques on the satellite's orbit.
For satellites in orbits exterior to the ring, these resonant ``Lindblad'' torques cause the satellite to migrate away from the ring \citep{Goldreich:1982cj}.

In addition to interacting with ring particles, ring satellites gravitationally interact with the primary body, which gives rise to tidal torques \citep{Murray:1999th}.
While inner Lindblad torques always cause outward satellite migration, the direction of the tidal torques depends on the relationship between the satellites' semi-major axes and the synchronous (or corotation) orbit.
If a satellite orbits beyond the synchronous orbit, tides will increase the satellite's semi-major axis, causing it to migrate away from the primary and the ring.
Tides cause the orbits of satellites that orbit within the synchronous orbit to decay, causing the satellite to migrate inward, towards the primary \citep{Murray:1999th}.

The formation of ring satellites from planetary rings, and the subsequent interactions within the ring-satellite environment indicate that these are strongly coupled systems.
Numerical simulations modeling the dynamics of ring-satellite systems demonstrate how coupled ring-satellite systems evolve over time.
Models of the spreading of a viscous ring that forms satellites which exchange angular momentum with the ring and the primary body have been applied to the formation of the inner satellites of the giant planets \citep{Charnoz2010,Charnoz2011,Crida2012,Salmon2017}.
HM17 used a 1-D mixed Eulerian/Lagrangian numerical model called ``RING-MOONS'' to examine how Phobos may have formed from a ring-satellite system in orbit around Mars.
This study and others revealed that the long-term evolution of a ring-satellite system in orbit around Mars is strikingly different than for a similar system in orbit around Saturn \citep{Rosenblatt2012,Rosenblatt2016,Hesselbrock2017,Charnoz2010,Charnoz2011}.
While Saturn's ring satellites always migrate away from the rings, ring satellites generated by a Martian ring can migrate inward where they can be tidally disrupted to form new rings.

In this work we investigate the major factors that control the long term evolution of coupled ring-satellite systems, including the location of the synchronous orbit relative to the location where ring satellites form.
Here we generalize the model in HM17 to other planetary systems.
In Section~\ref{sec:physics} we develop a model for the dynamics of coupled ring-satellite systems and show how various parameters affect the evolution of these systems.
We show that coupled ring-satellite systems can be classified into three distinct regimes of dynamical evolution, which we term ``Boomerang,'' ``Slingshot,'' and ``Torque-Dependent.''
In Section~\ref{sec:regimes} we explore the long-term evolution of ring-satellite systems in each of the three regime.
In Section~\ref{sec:SS} we show that representative examples of systems in all three evolution regimes can be found among bodies in our solar system.
We identify the inner Uranian ring-satellite system as a particularly interesting case.
In Section~\ref{sec:Uranus} we use our model to propose a formation hypothesis for the rings and inner satellites of Uranus, up to and including Miranda, from a massive ancient ring that may have formed from a giant impact onto Uranus.

\section{The Dynamics of Coupled ring-satellite Systems}\label{sec:physics}

In this section we describe the various dynamical processes that affect the evolution of ring-satellite systems over time.
Our goal is to identify parameters which strongly affect the long-term evolution of coupled ring-satellite systems.
The major processes in our model that affect the formation and orbital migration of rings and satellites includes the locations of the Roche limits, Lindblad torques, and tidal torques.
We define the Roche limits in Section~\ref{sec:roche}, the effect of Lindblad torques in Section~\ref{sec:lindblad}, and the effect of tidal torques in Section~\ref{sec:tides}.
We define three regimes for the evolution of coupled ring-satellite systems that depend on the rotational period of the primary and the bulk densities of the satellites and the primary.
The implications of these regimes on the dynamical evolution of satellites produced by the ring are discussed in Section~\ref{sec:regimes}.

\subsection{The Roche Limits}\label{sec:roche}

The Roche Limit is the semi-major axis at which the tensional forces acting on a cohesionless satellite due to tides are greater than the compressional forces from self-gravity \citep{Murray:1999th}.
It can be calculated by solving for the semi-major axis at which a particle on the equator of a synchronously rotating satellite has zero acceleration.
For orbits outside the Roche limit, the self-gravity of the satellite dominates and the particle remains on the body's surface.
Conversely, for orbits interior to the Roche limit, the gravitational attraction of the primary dominates and the particle is accelerated away from the satellite's surface.

The shape of the satellite plays a role in the definition of the Roche limit.
Two common definitions for the Roche limit are the ``rigid Roche limit'' (RRL) and the ``fluid Roche limit'' (FRL).
The assumption that the satellite can maintain a rigid spherical shape gives rise to the RRL.
The semi-major axis of the RRL is \citep{Murray:1999th}:
\begin{equation}\label{eqn:rrl}
a_{RRL} = R_{p}\left(\frac{3 \rho_{p}}{\rho_{s}}\right)^{1/3} \approx 1.442 R_{p}\left(\frac{\rho_{p}}{\rho_{s}}\right)^{1/3}.
\end{equation}
Here $R_{p}$ and $\rho_{p}$ are the radius and bulk density of the primary, while $\rho_{s}$ is the bulk density of the satellite.

Most satellites are not rigid spheres, and therefore the RRL is a lower limit for a ``cohesionless'' satellite.
At the opposite extreme, the satellite behaves as a fluid that can flow into a hydrostatic equilibrium shape.
The resulting ellipsoidal figure of equilibrium for a fluid satellite is a prolate spheroid \citep{Chandra69}.
Because the particles on the equator of a prolate body are located a greater distance from the body center as compared to a spherical body, the self-gravity component of the acceleration is reduced when calculating the ``fluid Roche limit'' (FRL) \citep{Chandra69}.
Thus, the FRL is farther from the primary than the RRL.
The semi-major axis of the FRL is \citep{Murray:1999th}:
\begin{equation}\label{eqn:frl}
a_{FRL} \approx 2.456 R_{p}\left(\frac{\rho_{p}}{\rho_{s}}\right)^{1/3}.
\end{equation} 

Although the material in planetary rings is solid, the interactions between ring particles can be approximated as a fluid.
This means that ring material orbiting inside the FRL is continually tidally disrupted, and is unable to accrete into satellites.
If ring material is transported beyond the FRL, where tidal forces are weaker, it can accrete into satellites.
Thus, the FRL marks the approximate outside boundary of a massive ring.

Once formed, interparticle forces within the satellite may prevent it from attaining a hydrostatic equilibrium shape.
Internal friction can allow even strengthless aggregates of particles to maintain non-hydrostatic shapes.
Therefore, the location where a fully formed satellite can be disrupted into a ring may be inward of where it formed.
The RRL is approximately the innermost boundary where such a strengthless satellite could hold together \citep{Hesselbrock2017}.

For a given primary body, calculating the Roche limits allows us to estimate where rings and satellites may orbit the body.
Ring material is confined to orbit inside the FRL.
While satellites may only accrete outside the FRL, if they have internal cohesion they may exist anywhere outside the RRL.
Assuming the satellites have no strength, no satellites should orbit inside the RRL, as they would be tidally disrupted \citep{Black2015}.

As the particles within rings collide and exchange energy and angular momentum, any ring in orbit around a primary will spread out overtime due to a process called ``viscous spreading'' \citep{Lynden-Bell1974,Goldreich:1982cj}.
This spreading process causes material to be transported both inwards, towards the primary, and outwards toward the FRL.
Thus, as the ring spreads out, some material is transported beyond the FRL where it may accrete into satellites \citep{Charnoz2010,Crida2012,Rosenblatt2012,Salmon2017,Hesselbrock2017}.

\subsection{Lindblad Torques}\label{sec:lindblad}

Once a satellite has formed, it begins to gravitationally interact with material in the ring.
The orbital speed of the interior ring material is greater than that of the exterior orbiting satellite.
Ring material is gravitationally attracted to the perturbing satellite, causing a density perturbation that is carried ahead of the satellite.
The interactions between the ring material and the satellite are strongest for material in a first order resonance with the satellite.
The resonant perturbations cause the ring material and the exterior satellite to exchange angular momentum through a series of torques called ``Lindblad torques'' \citep{Lynden-Bell1974,Goldreich:1979iq}.
Lindblad torques concentrate the ring material into spiral density waves \citep{Goldreich:1979iq} which in turn perturb the orbit of the exterior satellite.
The resonant interaction transfers angular momentum from the ring to the exterior satellite, increasing the satellite's semi-major axis.
While the satellite and the ring are gravitationally attracted, Lindblad torques work to repel the satellite from the ring \citep{Esposito2006}.

We may determine the locations within a ring that are in resonance with a satellite.
For a satellite in a first order resonance of order $\mathcal{M} (>1)$, the Lindblad resonance locations in a Keplerian ring interior to the satellite $r_{\mathcal{M}}$ can be found as \citep{Takeuchi96}:
\begin{equation}\label{eqn:r_L}
r_{\mathcal{M}} = \left(1 - \frac{1}{\mathcal{M}}\right)^{2/3}a,
\end{equation}
where $a$ is the semi-major axis of the exterior satellite.

The magnitude of the Lindblad torques are a function of the distance to the satellite and the surface-mass density of the ring near the resonance.
To first-order, the Lindblad torque exerted onto a satellite with semi-major axis $a$ by the ring at resonance location $r_{\mathcal{M}}$ is given as \citep{Tajeddine2017}:
\begin{equation}\label{eqn:Gamma_Mexplicit}
	\Gamma_{\mathcal{M}} = \mp \frac{4\pi^{2}}{3}\left(\frac{\mathcal{M}}{\mathcal{M} - 1}\right)\sigma(r_{\mathcal{M}})\left(r_{\mathcal{M}}^{2}\beta n \frac{M_{s}}{M_{p}} A_{\mathcal{M}}\right)^{2}.
\end{equation}
Here $M_{p}$ and $M_{s}$ are the mass of the primary and the satellite, and $\sigma(r_{\mathcal{M}})$ is the surface-mass density of the ring at the location of the Lindblad resonance.
The mean motion $n = \sqrt{GM_{p}/a^3}$, $\beta = r_{\mathcal{M}}/a$, and $G$ is Newton's gravitational constant.
$A_{\mathcal{M}}$ is a dimensionless quantity that depends upon Laplace coefficients and their derivatives.
It is calculated as:
\begin{equation}
	A_{\mathcal{M}} = \frac{1}{2}\left[2\mathcal{M}b_{1/2}^{\mathcal{M}}(\beta) + \beta Db_{1/2}^{\mathcal{M}}(\beta) \right].
\end{equation}
$b_{1/2}^{\mathcal{M}}(\beta)$ is the Laplace coefficient and $D$ is its derivative.
\cite{Murray:1999th} provides an algorithm to calculate the Laplace coefficients and their derivatives.
The derivative, $D$ is simply a function of Laplace coefficients.
\begin{equation}
	D b_{1/2}^{\mathcal{M}}(\beta) = \frac{1}{2}\left[b_{3/2}^{\mathcal{M}-1}(\beta) - 2\beta b_{3/2}^{\mathcal{M}}(\beta) + b_{3/2}^{\mathcal{M} + 1}(\beta) \right].
\end{equation}

For a satellite exterior to a ring the sign of the torque is positive, while for a satellite interior to a ring the sign of the torque is negative.
We will only consider the case of satellites exterior to rings here.
The total torque $\Gamma$ exerted onto the satellite by the ring is the sum of the individual resonant torques, or:
\begin{equation}\label{eqn:total_torque}
\Gamma = \sum_{\mathcal{M}=2}^{\infty} \Gamma_{\mathcal{M}}.
\end{equation}
Lindblad torques transfer orbital angular momentum from the ring particles to the satellites which causes the satellites to migrate away from the ring.
The change in an exterior satellite's semi-major axis due to an inner Lindblad torque is \citep{Takeuchi96}:
\begin{equation}\label{eqn:lindblad}
\frac{da}{dt} = \frac{2\Gamma}{M_{s}}\left(\frac{a}{GM_{p}}\right)^{1/2}.
\end{equation}

The ring can exert a Lindblad torque on the satellite as long as some ring material is in a first-order resonance with the satellite.
Therefore, there is a maximum distance a satellite can be perturbed to via Lindblad torques, as eventually the satellite will orbit too far away to be in resonance with any material in the ring.
The farthest the satellite may be perturbed to via Lindblad torques is when the satellite is in resonance with the material at the edge of the ring, at the FRL.
To find the satellite semi-major axes that may be in resonance with the ring edge, we substitute $a_{FRL}$ for $r_{\mathcal{M}}$ and rearrange Equation~\ref{eqn:r_L} to yield:
\begin{equation}
a_{\mathcal{M}} = a_{FRL}\left(1 - \frac{1}{\mathcal{M}}\right)^{-2/3}.
\end{equation}
The greatest satellite semi-major axis that the ring edge may be in resonance with occurs when $\mathcal{M} = 2$, which corresponds to the $2:1$ mean motion resonance.
Therefore, we find that the maximum orbit a satellite may migrate to via Lindblad torques is:
\begin{equation}\label{eqn:a_max}
a_{Lind} = 4^{1/3}a_{FRL} \approx 2.456 R_{p}\left(\frac{4\rho_{p}}{\rho_{s}}\right)^{1/3}.
\end{equation}

\subsection{Tidal Torques}\label{sec:tides}

In addition to accelerating the material in the ring, the satellite also exerts an acceleration on the surface of the primary \citep{Murray:1999th}.
The acceleration from the orbiting satellite varies in magnitude across the primary's surface and is greatest for the surface closest to the satellite.
The gradient tidal potential across the primary's surface distorts the primary's shape, creating a tidal bulge.
The internal structure of the primary determines the response of the surface to the tidal potential.
Internal friction dissipates the tidal acceleration and results in a lag between the tidal disturbance and the tidal response.
The effect of tidal dissipation can lead to dramatic physical and orbital consequences for the primary and the satellite.

Although the satellite creates the tidal bulge on the primary, the acceleration of the bulge on the satellite exerts a torque on the satellite.
The consequence of the tidal torque is dependent upon the semi-major axis of the satellite relative to the synchronous orbit.
For a satellite in a Keplerian orbit, the synchronous orbit can be calculated as:
\begin{equation}\label{eqn:synch}
a_{synch} = \left[\frac{G(M_{p}+M_{s})T_{p}^{2}}{4\pi^{2}}\right]^{1/3},
\end{equation}
where $T_{p}$ is the rotational period of the primary.
If the satellite is in a synchronous orbit with the primary, it completes one orbit for every full revolution of the primary, and therefore is always aligned with the tidal bulge.
The gravitational attraction between the tidal bulge and the satellite is perpendicular to the satellite's motion, and no torque results.

However, if the satellite orbits interior to the synchronous orbit, its orbital period is shorter than the rotational period of the primary.
As the satellite orbits, it passes over the surface of the primary.
Due to the lag in the formation of the tidal bulge, the satellite is always ``ahead'' of the tidal bulge on the primary surface.
The gravitational attraction between the tidal bulge and the satellite is no longer strictly perpendicular to the satellite's motion, and the tidal bulge exerts a torque on the satellite.
For satellites orbiting interior to the synchronous orbit this torque transfers angular momentum from the satellite's orbit to the spin angular momentum of the primary, causing the satellite to migrate towards the primary.

Alternatively, if the satellite orbits exterior to the synchronous orbit, its orbital period is longer than the rotational period of the primary.
As the satellite orbits, the surface of the primary rotates past the satellite.
Due to the lag in the formation of the tidal bulge, the satellite is always ``behind'' the tidal bulge on the primary surface.
The bulge again exerts a torque on the satellite.
However, for satellites orbiting exterior to the synchronous orbit the tidal torque transfers spin angular momentum from the primary to the satellite's orbital angular momentum, causing the satellite to migrate away from the primary.
Thus, satellites which lie interior to the synchronous orbit migrate inwards by tides, whereas satellites which orbit exterior migrate outwards.

For satellites that are not located at $a_{synch}$, the transfer between the satellite's orbital angular momentum and the spin angular momentum of the primary causes the satellite to migrate in its orbit.
The torque exerted by the tidal bulge changes the semi-major axis of the satellite.
The change in the satellite's semi-major axis due to the tidal interaction is calculated as \citep{Murray:1999th}:
\begin{equation}\label{eqn:tide}
	\frac{da}{dt} = \mp \frac{3nM_{s}R_{p}^{5}}{a^{4}M_{p}} \left(\frac{k_{2}}{Q}\right)\left[1+\frac{51e^{2}}{4}\right].
\end{equation}
Here $e$ is the eccentricity of the satellite while $k_{2}$ and $Q$ are the tidal potential love number and tidal quality factor of the primary.
The upper ($-$) sign describes a satellite with $a < a_{synch}$, and the lower ($+$) sign describes a satellite with $a > a_{synch}$.

\subsection{Satellite Migration}

The tidal interaction with the primary, as well as the Lindblad torques combine to drive the orbital migration of the satellite.
To determine the change in the semi-major axis of a satellite orbiting exterior to a ring we may combine Equations \ref{eqn:lindblad} and \ref{eqn:tide}.
The total change in the semi-major axis of a satellite in time due to Lindblad and tidal torques is given as \citep{Rosenblatt2012}:
\begin{equation}\label{eqn:dadt}
	\frac{da}{dt} = \mp \frac{3nM_{s}R_{p}^{5}}{a^{4}M_{p}} \left(\frac{k_{2}}{Q}\right)\left[1+\frac{51e^{2}}{4}\right] + \frac{2\Gamma}{M_{s}}\left(\frac{a}{GM_{p}}\right)^{1/2}.
\end{equation}

As shown in Equation~\ref{eqn:dadt}, if $a > a_{synch}$ for a satellite exterior to the ring, both terms are positive, $da/dt$ is positive, and the semi-major axis of the satellite increases.
However, if $a < a_{synch}$, the two terms are opposite in sign, and a competition exists:  the tidal torques work to evolve the satellite inwards while the Lindblad torques work to drive the satellite away.
$da/dt$ is no longer strictly positive because the relative magnitude of these two torques determines the sign of Equation~\ref{eqn:dadt}.
The location of the synchronous orbit relative to the FRL determines whether tidal torques cause newly formed ring satellites to migrate towards or away from the ring, thus playing a key role in understanding the migration of satellites accreting at the ring edge.

\begin{figure*}
\gridline{\fig{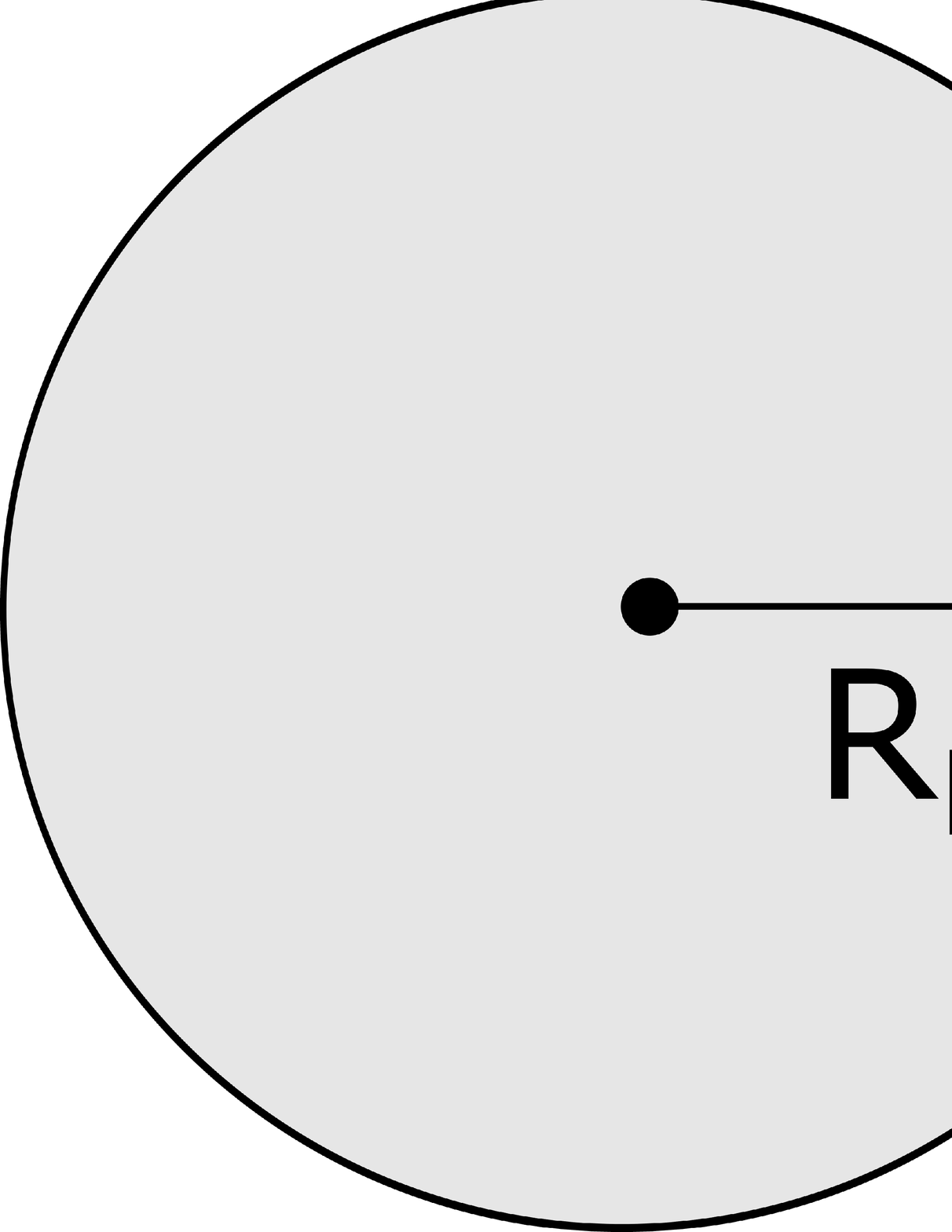}{0.6\textwidth}{(a) Boomerang}
          }
\gridline{\fig{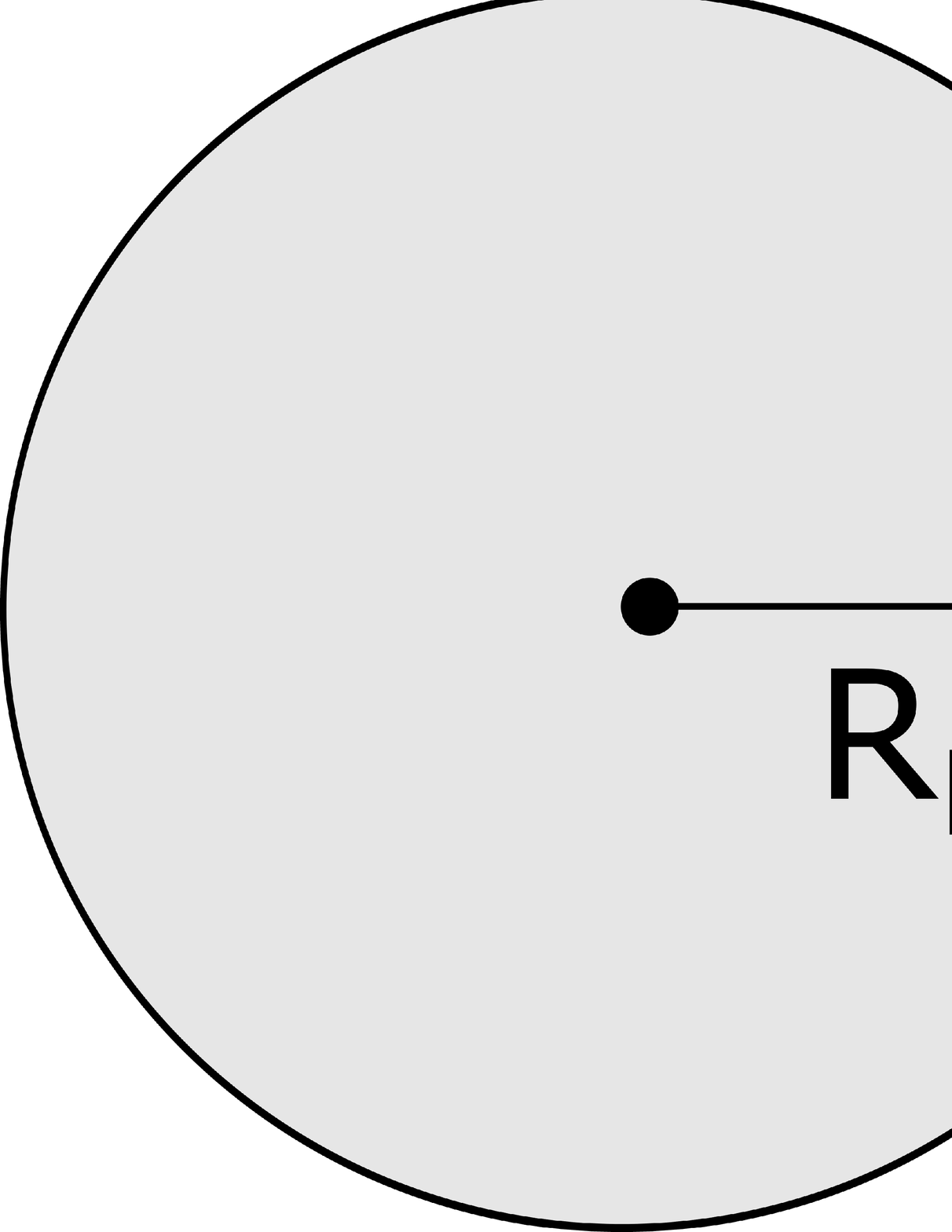}{0.6\textwidth}{(b) Torque-Dependent}
          }
\gridline{\fig{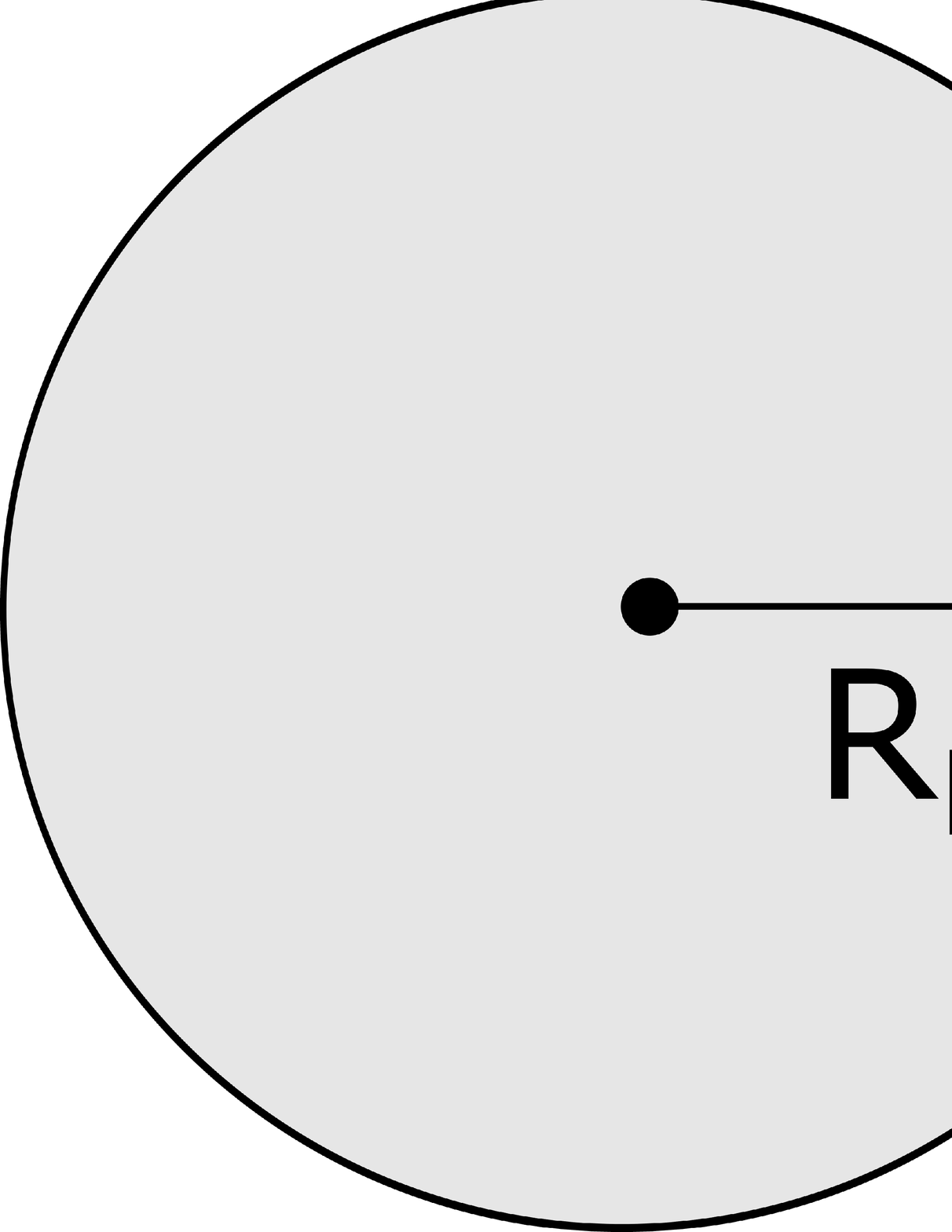}{0.6\textwidth}{(c) Slingshot}}
\caption{Diagram showing the relationships that define our three evolution regimes for coupled ring-satellites.  The large grey circle represents a primary body of radius $R_{p}$ orbited by a ring (shaded region), and a small satellite (small circle) that has formed at the ring edge.  The rigid Roche limit (RRL), fluid Roche limit (FRL), the maximum orbit Lindblad torques may migrate a satellite ($a_{Lind}$), and the synchronous orbit ($a_{synch}$) are all marked with vertical lines.  Distances are shown in units of primary radii.  The primary and satellite are assumed to have the same density.  (a) Boomerang regime:  For a slowly rotating primary, $a_{synch}$ lies beyond $a_{Lind}$.  (b) Torque-Dependent regime: For a moderate rotation period, $a_{synch}$ lies between the FRL and $a_{Lind}$.  (c) Slingshot regime:  For rapidly rotating primaries $a_{synch}$ lies inside the FRL.\label{fig:setup}}
\end{figure*}

The locations of the RRL, FRL, and $a_{Lind}$ are all functions of the bulk density of the primary body and the satellites.
Independently, the synchronous orbit is only a function of the rotational period of the primary.
From Equation~\ref{eqn:synch} we see that the rotation period of the primary and the location of the synchronous orbit are directly related.
In Figure~\ref{fig:setup} we display three identical ring-satellite systems, however in each panel the rotation period of the primary has been varied.
In each panel, $\rho_{s}/\rho_{p} = 1$.
Using Equations \ref{eqn:rrl}, \ref{eqn:frl}, and \ref{eqn:a_max} we may determine the locations of the RRL, FRL, and $a_{Lind}$ in terms of primary radii.
For a given primary rotation period, we may determine the location of $a_{synch}$.
In Figure~\ref{fig:setup}a we have marked the location of $a_{synch}$ for a primary with $T_{p}$ such that $a_{synch} > a_{Lind}$.
In Figure~\ref{fig:setup}b we have displayed a similar system, however the primary rotates with a shorter $T_{p}$ such that $a_{FRL} < a_{synch} < a_{Lind}$.
Finally, in Figure~\ref{fig:setup}c, we display an identical system, but with a rapidly rotating primary such that $a_{synch} < a_{FRL}$.

We define a function $f$, which we use to determine the location of $a_{synch}$ relative to $a_{FRL}$ and $a_{Lind}$.
The value of $f$ for a coupled ring-satellite system determines the regimes shown in Figure \ref{fig:setup}.
To define $f$, we derive the conditions for $a_{synch} = a_{Lind}$.
We set $a_{synch} = a_{Lind}$ using Equations \ref{eqn:synch} and \ref{eqn:a_max}:
\begin{equation}\label{eqn:f_derive}
	 \frac{1}{2.456^{3}} \frac{G\rho_{s}T_{p}^{2}}{3 \pi}\left(1 + \frac{M_{s}}{M_{p}}\right) = 4.
\end{equation}
We define the left hand side of Equation~\ref{eqn:f_derive} to be the function $f$, such that:
\begin{equation}\label{eqn:f}
f(T_{p}, \rho_{s}) = \frac{1}{2.456^{3}} \frac{G\rho_{s}T_{p}^{2}}{3 \pi}\left(1 + \frac{M_{s}}{M_{p}}\right).
\end{equation}
Thus, the right hand side of Equation~\ref{eqn:f_derive} yields the condition when $a_{synch} = a_{Lind}$.
In this scenario $f(T_{p}, \rho_{s}) = 4$.
Therefore, for ring-satellite systems where $f(T_{p}, \rho_{s}) > 4$, the synchronous orbit lies beyond the maximum orbit to which a ring-accreted satellite could migrate via Lindblad torques.

In order to determine the location of $a_{synch}$ relative to $a_{FRL}$ we may derive the conditions for when $a_{synch} = a_{FRL}$.
Using Equations \ref{eqn:synch} and \ref{eqn:frl}:
\begin{equation}
\frac{1}{2.456^{3}}\frac{G\rho_{s}T_{p}^{2}}{3\pi}\left(1 + \frac{M_{s}}{M_{p}}\right) = 1.
\end{equation}
During this derivation we recover the function $f$ and the conditions for when $a_{synch} = a_{FRL}$.
In this scenario $f(T_{p}, \rho_{s}) = 1$.

The function $f(T_{p}, \rho_{s})$ may be applied to any planetary ring system to determine in which formation regime the system exists.
In ring systems where $f(T_{p}, \rho_{s}) > 4$, the synchronous orbit lies beyond the maximum orbit a satellite could migrate via Lindblad torques (i.e. Figure~\ref{fig:setup}a).
For ring systems where $1 < f(T_{p}, \rho_{s}) < 4$, the synchronous orbit lies between the FRL and $a_{Lind}$ (i.e. Figure~\ref{fig:setup}b).
Finally, in ring systems where $f(T_{p}, \rho_{s}) < 1$, the synchronous orbit lies inside the FRL (i.e. Figure~\ref{fig:setup}c).
In each scenario depicted in Figure~\ref{fig:setup}, and further defined by Equation~\ref{eqn:f}, the orbital migration of the satellite that has accreted at the ring edge, as determined by Equation~\ref{eqn:dadt}, is different.
In Section~\ref{sec:regimes}, we explore the long-term evolution of ring-satellite systems in each of the three formation regimes.

\section{Evolution of Ring Accreted Satellites}\label{sec:regimes}

The scenarios depicted in Figure~{\ref{fig:setup}} represent three distinct formation regimes for satellites accreting from planetary rings
In the ``Boomerang'' regime the synchronous orbit lies beyond the maximum orbit a satellite may be perturbed via Lindblad torques, as shown for the system in Figure~{\ref{fig:setup}}a.
In the ``Slingshot'' regime the synchronous orbit lies inside the ring edge, as displayed in Figure~\ref{fig:setup}c.
And finally, in the ``Torque-Dependent'' regime the synchronous orbit lies between the edge of the ring and $a_{Lind}$, as in Figure~\ref{fig:setup}b, creating a competition between the tidal and Lindblad torques exerted on a newly formed satellite.
Depending upon the magnitudes of the ring and tidal torques satellites in this regime either migrate away from the primary and survive, similar to satellites in a Slingshot regime, or migrate towards the RRL, similar to satellites in a Boomerang regime.

In Section~\ref{sec:physics} we defined the function $f(T_{p}, \rho_{s})$, which we use to determine the location of $a_{synch}$ relative to $a_{Lind}$ and the FRL.
This relation can be applied to any known or hypothesized planetary ring system in order to predict the long term evolution of satellites that may accrete at the FRL.
In the following section we will use this relation to characterize and analyze each of the regimes for satellite evolution.

\subsection{The ``Boomerang'' Regime:  $a_{Lind} < a_{synch}$}\label{sec:boomerang}

For slowly rotating primaries, the synchronous orbit lies far from the primary's surface.
For planetary ring systems where $f(T_{p}, \rho_{s}) > 4$, $a_{synch}$ lies beyond $a_{Lind}$, as in Figure~\ref{fig:setup}a.
We may examine Equation~\ref{eqn:dadt} to determine the orbital migration of a satellite that accretes from a planetary ring in the Boomerang regime.
In the Boomerang regime $da/dt$ for a newly accreted satellite at the FRL is not necessarily positive.
The ring may have sufficient mass for Lindblad torques to temporarily drive a satellite away from the primary.
However, planetary tides always work to migrate the satellite inwards.

As the satellite is driven away from the ring, eventually an equilibrium between the Lindblad and tidal torques is reached, and $da/dt = 0$.
Over time the ring is depleted of material as mass is deposited onto the primary through the ring's inner edge, and mass is lost to satellite formation at the ring's outer edge \citep{Rosenblatt2012,Rosenblatt2016,Hesselbrock2017}.
As the ring loses mass, the effect of Lindblad torques on the satellites' evolution is diminished until tidal torques with the primary dominate.
Similar to a boomerang, satellites in this regime may initially be driven away from the ring, but eventually are driven inwards as Equation~\ref{eqn:dadt} becomes negative.
Regardless of how massive the ring is, the orbit of a satellite accreting at the FRL in this ``boomerang'' regime will eventually decay.
Therefore, we define the Boomerang regime as any planetary ring system with $f(T_{p}, \rho_{s}) > 4$.

Furthermore, as the satellite approaches the RRL, the magnitude of the tidal stress increases and material begins to leave the surface of the satellite.
This removed material creates a collisional cascade that quickly disrupts the body.
The disrupted satellite material then forms a new ring of material that would begin to deposit its material onto the primary's surface \citep{Black2015}.
As the new ring begins to spread out through viscous spreading, it may transport material beyond the FRL to form a new generation of satellites \citep{Hesselbrock2017}.
However, as these satellites accrete in the Boomerang regime they will follow a similar migration path.
We recently showed that the innermost satellite of Mars, Phobos, may have formed in the Boomerang regime \citep{Hesselbrock2017}.
Therefore, slowly rotating primaries where $a_{synch}$ lies beyond $a_{Lind}$ serve as desirable candidates for ring-satellite cycles, as proposed in HM17.

\subsection{The ``Slingshot'' Regime:  $a_{synch} < a_{FRL}$}\label{sec:non_boom}

As shown in Section~\ref{sec:physics}, for ring systems where $f(T_{p}, \rho_{s}) < 1$ the edge of the ring lies beyond the synchronous orbit.
For a satellite accreting at the edge of the ring in Figure~\ref{fig:setup}c, both the tidal torques and the Lindblad torques work to increase the semi-major axis of the satellite.
$da/dt$ in Equation~\ref{eqn:dadt} is strictly positive and any satellites that form at the ring edge are driven away from the primary.
Similar to a projectile fired from a slingshot, in these systems satellites that form at the FRL are forever driven away from the primary.
We define the ``Slingshot'' regime as any planetary ring system where $f(T_{p}, \rho_{s}) < 1$.

\subsection{The ``Torque-Dependent'' Regime:  $a_{FRL} < a_{synch} < a_{Lind}$}\label{sec:mass_dep}

Similar to the Boomerang regime, for a ring-satellite system where $a_{synch}$ lies between the FRL and $a_{Lind}$, satellites that accrete at the FRL experience a competition between Lindblad and tidal torques.
As the satellites are pulled inwards by planetary tides, $da/dt$ in Equation~\ref{eqn:dadt} is not strictly positive.
However, if the magnitude of the total Lindblad torque is greater than the inward tidal torque, it may be possible for Equation~\ref{eqn:dadt} to remain positive for a sufficient amount of time for a satellite to migrate beyond $a_{synch}$.
At this point in the satellite's migration Equation~\ref{eqn:dadt} becomes strictly positive, regardless of the magnitude of the Lindblad torque, and the satellite survives to migrate away from the primary, similar to the evolution of a satellite in the Slingshot regime.

Alternatively, if the magnitude of the tidal torque is greater than the total Lindblad torque for a satellite orbiting inside $a_{synch}$, Equation~\ref{eqn:dadt} is negative and the satellite would be unable to migrate beyond $a_{synch}$.
As $da/dt < 0$, the satellite's orbit migrates inwards towards the RRL.
Depending on the magnitude of the Lindblad torques, satellites that accrete in this regime may undergo an identical evolution to satellites that accrete in the Boomerang regime.
As the outcome of the satellites' orbital migration is dependent upon the competition between the tidal and Lindblad torques, we refer to this regime as the ``Torque-Dependent'' regime and define the regime as any ring system where $1 < f(T_{p}, \rho_{s}) < 4$.

In order to predict the migration of satellites accreting in the Torque-Dependent regime we must closely examine the contribution of the Lindblad and tidal torques to the satellite's migration.
The magnitude of the total Lindblad torque in Equation~\ref{eqn:total_torque} compared to the tidal torque with the primary is a determining factor to the migration of a satellite interior to $a_{synch}$.
Examining Equations \ref{eqn:dadt} and \ref{eqn:total_torque}, we see that depending on the tidal parameters of the bodies and the surface-mass density of the ring, it is possible for the total Lindblad torque exerted on the satellite to be greater in magnitude than the tidal torque.
In this case, $da/dt$ in Equation~\ref{eqn:dadt} is positive and the satellite is driven away from the ring edge.
We define this as a scenario in which the Lindblad torques are able to overcome the tidal interactions that work to drive the satellite inwards.

We may also examine when the magnitude of the total Lindblad torque is less than that of the tidal torque for a satellite interior to $a_{synch}$.
In these cases, $da/dt$ in Equation~\ref{eqn:dadt} is negative, and the Lindblad torques are unable to overcome the tidal interactions.
Thus, if the surface-mass density of the ring is small, and/or the satellite is strongly affected by tidal interactions with the primary, the satellite will migrate inwards.

It is possible to determine whether the Lindblad torques dominate over tides or not by setting Equation~\ref{eqn:dadt} equal to zero.
This allows us to determine the conditions for which the Lindblad and tidal torques are equal in magnitude.
If the magnitude of the tidal term in Equation~\ref{eqn:dadt} is less than the magnitude of the Lindblad term, the satellite will be driven outwards.
By substituting the total Lindblad torque in Equation~\ref{eqn:total_torque} into the Lindblad term in Equation~\ref{eqn:dadt}, we find that the satellite will be driven away from the ring so long as the following statement is true:
\begin{equation}\label{eqn:threshold}
	\sum_{\mathcal{M} = 2}^{\infty}\left(\frac{\mathcal{M}}{\mathcal{M} - 1}\right)\sigma (r_{\mathcal{M}})r_{\mathcal{M}}^{3} A_{\mathcal{M}}^{2} > \frac{9}{8\pi^{2}}\left(\frac{k_{2}}{Q}\right)\frac{M_{p}R_{p}^{5}}{a^{4}}.
\end{equation}
Here we have assumed that the satellite eccentricity is zero.
Equation~\ref{eqn:threshold} is similar to Equation~A.2 in \citet{Rosenblatt2012}.

If Equation~\ref{eqn:threshold} is true for the entirety of the satellite's evolution to the synchronous orbit, the satellite will be driven away from the primary until it reaches an orbit beyond $a_{synch}$.
At this point $da/dt$ becomes strictly positive and the satellite will be forever driven away from the primary.
If the ring is sufficiently massive the satellite will follow an evolution similar to satellites in a Slingshot regime.
However, as discussed in Section~\ref{sec:boomerang}, material is constantly removed from the ring over time, diminishing the magnitude of the Lindblad torques.
If Equation~\ref{eqn:threshold} becomes false before the satellite has evolved to $a_{synch}$, the satellite will be pulled inwards towards the primary.
Similar to satellites in the Boomerang regime, the satellite's orbit will decay until either the satellite is tidally disrupted, forming a new ring, or the satellite is deposited onto the primary.

\section{A Look at Our Solar System}\label{sec:SS}

\begin{figure}[t]
\plotone{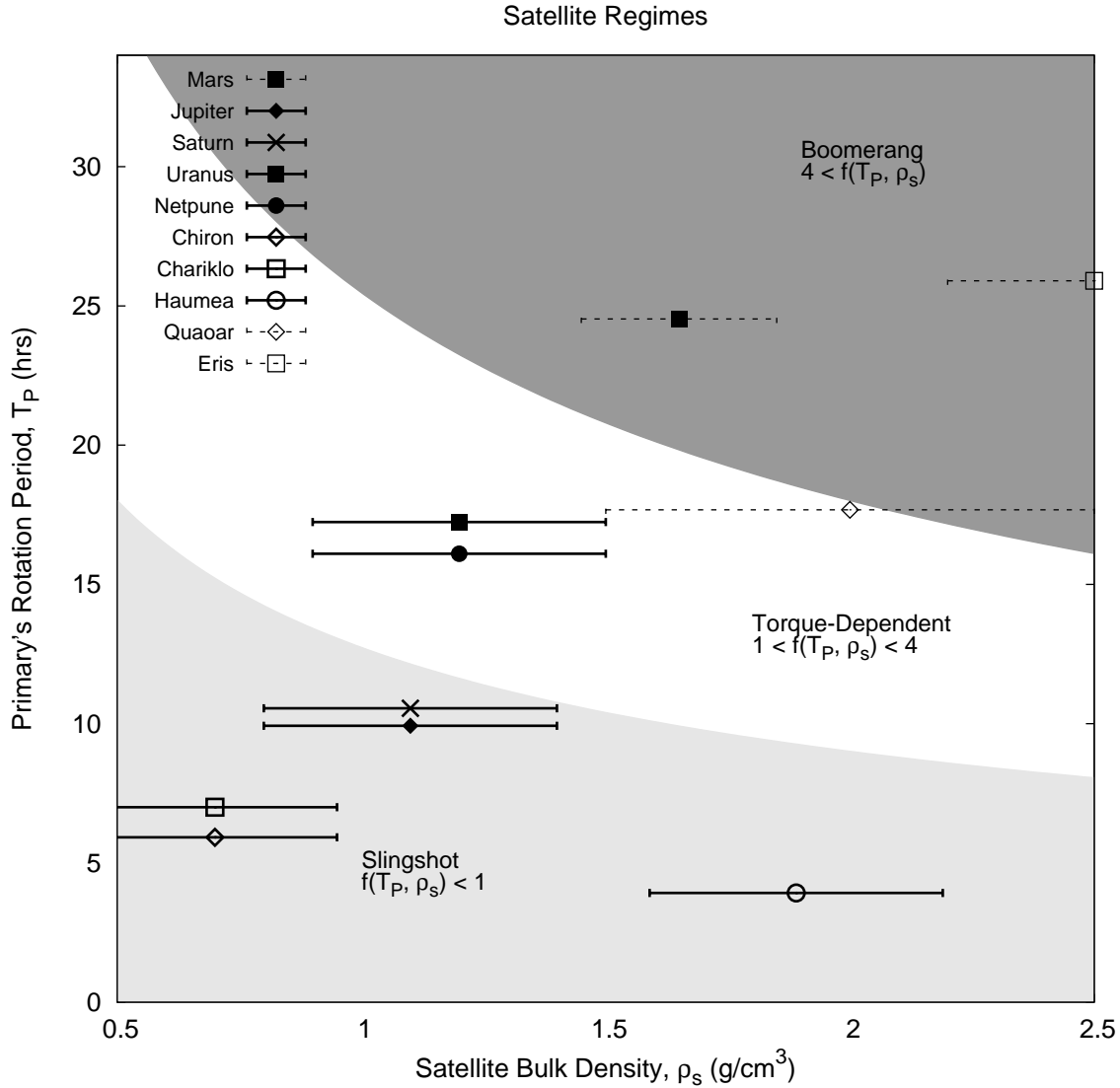}
\caption{The boundaries of our three ring-satellite evolution regimes defined by Equation~\ref{eqn:f} (assuming $M_{s}/M_{p} << 1$).  The dark gray regime marks systems where the synchronous orbit lies beyond the maximum orbit Lindblad torques may migrate a satellite (Boomerang regime), while the light gray regime marks systems where the synchronous orbit lies inside the edge of the ring (Slingshot regime).  The middle zone in white marks systems where the synchronous orbit lies between the edge of the ring and the maximum orbit (Torque-Dependent regime). Additionally, we display the expected regime for Roche-interior rings orbiting various bodies in the solar system given their rotation rates today and estimated satellite densities.  In this figure, $M_{s}/M_{p} << 1$.  Fast rotating primaries with low satellite densities fall within the Slingshot regime (e.g. Jupiter, Saturn), while slowly rotating primaries with high density satellites exist within the Boomerang regime (e.g. Mars).  The evolution of satellites accreting from a Roche-interior ring orbiting Uranus and Neptune is dependent upon the magnitude of the Lindblad and tidal torques.\label{fig:boomerang_ss}}
\end{figure}

In Section~\ref{sec:regimes}, we identified three regimes for the evolution of coupled ring-satellite systems.
We defined these regimes with the function $f$ in Equation~\ref{eqn:f}, which is dependent upon the bulk density of the satellite and the rotation period of the primary.
Figure~\ref{fig:boomerang_ss} plots the boundaries of the three regimes defined by Equation~\ref{eqn:f} assuming the satellite mass is small relative to the primary ($M_{s}/M_{p} << 1$).

The dark gray region in Figure~\ref{fig:boomerang_ss} corresponds to the Boomerang regime (see Figure~\ref{fig:setup}).
For slowly rotating primaries ($T_{p} \sim 25$ hrs) we find that the synchronous orbit lies beyond the maximum orbit for a wide-range of satellite densities.
The light gray region in Figure~\ref{fig:boomerang_ss} corresponds to the Slingshot regime.
For rapidly rotating primaries ($T_{p} \sim 10$ hrs) we find that the synchronous orbit lies inside the ring edge for a wide range of satellite densities.
Lastly, the white region in Figure~\ref{fig:boomerang_ss} corresponds to the Torque-Dependent regime.
For primaries with moderate rotation rates ($T_{p} \sim 15$ hrs), satellites with a wide range of densities accreting at the ring edge will undergo a competition between Lindblad and tidal torques.
If the magnitude of the Lindblad torque is greater than the tidal torques, the system will evolve similar to a Slingshot system.
Otherwise, the system will exhibit an evolution similar to a Boomerang system.

In Figure~\ref{fig:boomerang_ss} we use the current rotation period of primary bodies in the Solar System with estimates for the density of their secondaries to determine the evolution regime for real and hypothetical ring systems in the Solar System.
We see that the hypothetical Mars system, modeled in HM17, with its slow rotation rate and satellites with relatively high bulk densities is located within the Boomerang regime.
Additionally, we see that the Saturn system, with its rapid rotation rate and low-density satellites is located within the Slingshot regime.
This is in agreement with observations of the architecture of the inner ring satellites of Saturn and recent simulation results \citep{Charnoz2010,Charnoz2011,Salmon2017}.

Recently stellar occultations have revealed rings in orbit around small bodies.
These observations have shown that the centaurs ``10199 Chariklo'' and ``2060 Chiron'' are both orbited by ring systems \citep{Braga2014,Ortiz2015}.
Furthermore, \citet{Ortiz2017} also reports that a planetary ring orbits Haumea, a dwarf planet in the Kuiper belt with two known satellites.
All of these bodies are rapid rotators, with Haumea approaching a Jacobi ellipsoid in hydrostatic equilibrium.
As shown in Figure~\ref{fig:boomerang_ss}, the short spin period of these three bodies place them in the Slingshot regime for a wide range of satellite densities.
Satellites generated from these rings would migrate away from the primary body.
Therefore, it is unlikely the tidal disruption of an inwardly migrating satellite would have produced the ring systems observed today.

Uranus and Neptune both fall into the Torque-Dependent regime, as shown in Figure~\ref{fig:boomerang_ss}.
Both of these giant planets have rings and inner satellites that orbit near the rings' edges.
\cite{Leinhardt2012Disrupt} proposed a tidal disruption origin of the inner satellites of Uranus and Neptune.
Furthermore, some of these satellites lie within the synchronous orbit, but some lie beyond.
It may be possible that these planets had primordial rings that were sufficiently massive to cause satellites to migrate far from the FRL and beyond $a_{synch}$, similar to a system in the Slingshot regime.
However, as the primordial ring decayed over time the evolution of satellites may have transitioned to follow an orbital migration similar to a system in the Boomerang regime \citep{Charnoz2018}.

Unfortunately, the dynamics of the Neptune-Triton capture event makes it difficult to compare the locations of the satellites today to a primordial ring system.
Such a dynamic capture event is not thought to have occurred at Uranus, making a comparison between its present day satellite system to a primordial ring more straightforward.
As the ring-satellite system at Uranus lies within the Torque-Dependent regime, it serves as a candidate for special scrutiny \citep{Charnoz2018}.
By examining Equation~\ref{eqn:threshold}, we may hypothesize the type of massive primordial ring that may have produced the Uranian satellites and migrated them to their current orbits.
In the following section we will further examine the dynamics of the Uranian ring-satellite system in order to examine whether a massive primordial ring may have produced the system we observe today.

\subsection{A Torque-Dependent Regime Evolution for the Inner Uranian Satellites}\label{sec:Uranus}

\begin{figure}[t]
\begin{centering}
\includegraphics[width=3.5in]{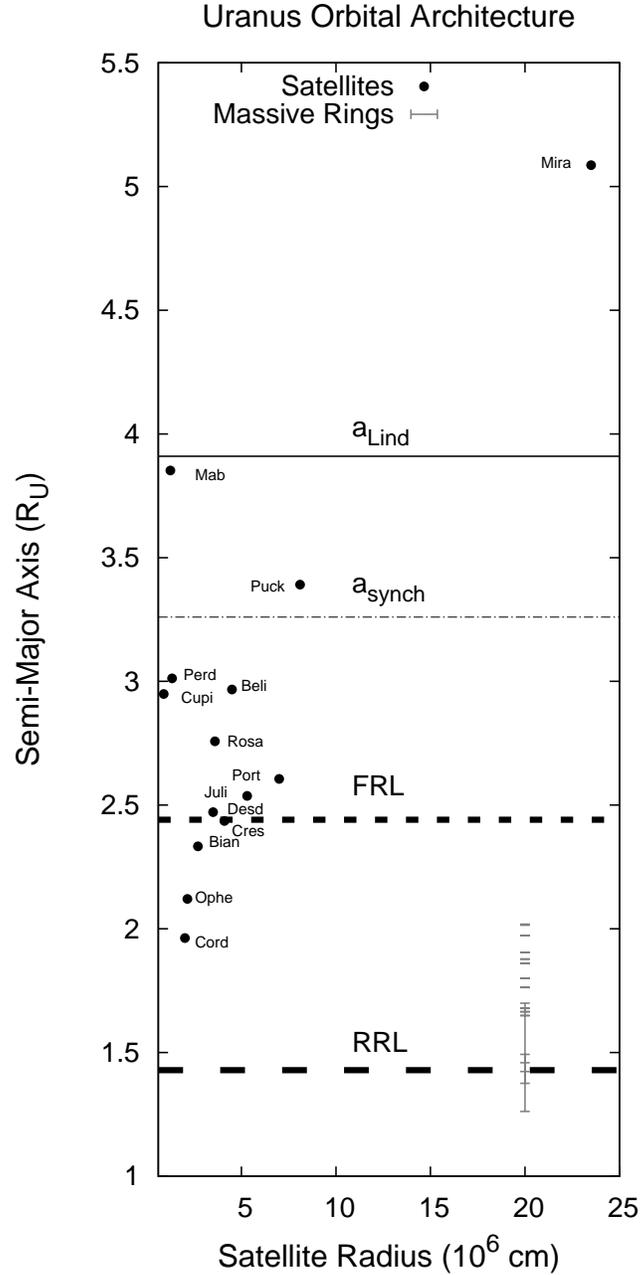}
\caption{Orbital architecture of the Uranian ring-satellite system today. The dark circles show the present-day orbits of known satellites (abbreviated names adjacent) while the gray lines mark the locations of known rings.  The location of the Rigid Roche Limit (RRL), Fluid Roche Limit (FRL), synchronous orbit ($a_{synch}$), and maximum orbit ($a_{Lind}$) for satellites perturbed by Lindblad Torques are displayed as horizontal lines.  Uranus has 13 satellites with orbits within $a_{Lind}$, indicating they may have accreted from an ancient Uranian ring.  The satellite Miranda, orbiting beyond $a_{Lind}$ may have also accreted from this ring.\label{fig:architecture}}
\end{centering}
\end{figure}

Examination of Figure~\ref{fig:boomerang_ss} reveals that Uranus falls within the Torque-Dependent regime for a wide range of satellite densities.
The planet is orbited by several rings and has multiple satellites, with $13$ known to orbit within $a_{Lind}$ (see Figure~\ref{fig:architecture}).
The masses and densities of $11$ of these inner satellites are not well constrained.
Each of these satellites are assumed to have a bulk density of $\sim 1.3 \text{ g cm}^{-3}$.
However, \citet{Jacobson92} used Voyager radar and image observations to determine the bulk density of Miranda to be $\sim 1.2 \text{ g cm}^{-3}$.
Additionally, \citet{Chancia2017} used perturbations in Uranus's $\eta$ ring to determine the bulk density of Cressida to be $\sim 0.86 \text{ g cm}^{-3}$.
Therefore, the other $11$ inner satellites of Uranus may have a bulk density less than typically assumed.

In this section, we will apply our coupled ring-satellite evolution model to gain insight into a possible formation and evolution scenario for the inner Uranian ring-satellite system.
Developed for HM17, RING-MOONS simulates the evolution of ring-satellite systems and is similar to the HYDRO-RINGS model used in \cite{Charnoz2011}.
In RING-MOONS, the grid-space of the ring is modeled as a series of Eulerian bins extending from the surface of the primary to beyond the FRL.
Following \cite{Salmon2010}, the material in the ring is modeled as a collection of uniform spherical particles that spread with variable viscosity.
Particles which spread just beyond the FRL accrete into Lagrangian satellites that accrete ring material within a feeding zone defined by the satellite's hill sphere.
Additionally, satellites exchange angular momentum with the primary, and with the material within the ring.
RING-MOONS calculates the migration of satellites by calculating the total ring torque exerted on the satellite via Equation \ref{eqn:total_torque} before numerically integrating Equation {\ref{eqn:dadt}}.
Satellites that approach each other are merged when their semi-major axes intersect their mutual Hill sphere.

In this section we implement RING-MOONS to examine whether Miranda may have formed from an ancient primordial ring in orbit around Uranus.
Uranus has a mass of $8.68 \times 10^{28}$~g, an average radius $R_{U} \sim 2.54 \times 10^{9}$~cm, a bulk density of $1.27 \text{ g cm}^{-3}$, and an obliquity of $97.8^{\circ}$.
A giant impact has been proposed to explain the large tilt of Uranus's rotation axis \citep{Slattery92}.
Such an impact may have occurred soon after the planet formed and may have placed a large amount of material into orbit.
This material would have collapsed into a Roche-interior ring around Uranus \citep{Morbidelli2012}.
The current inner satellite system may have accreted from this Roche-interior ring \citep{Charnoz2018}.
Furthermore, such an impact event may have produced Uranus's current rotation period of $17.24$ hours \citep{Slattery92}.

Miranda is a massive satellite that currently orbits Uranus beyond $a_{Lind}$.
It has a radius of $2.36 \times 10^{7}$~cm, an estimated mass of $6.6 \times 10^{22}$~g, and a bulk density of $1.2 \text{ g cm}^{-3}$ \citep{Jacobson92}.
Miranda has a semi-major axis of $1.30 \times 10^{10} \text{ cm} \sim 5.13 R_{U}$, and an eccentricity of $0.0013$.
The low eccentricity of Miranda's orbit is indicative of formation from a planetary ring \citep{Charnoz2010}.
Yet, Miranda orbits Uranus with an inclination of $4.3^{\circ}$, which is unexpected for satellites accreting from a planetary ring.
Miranda does not exhibit any mean motion resonances today, and the relatively high inclination of its orbit (as compared to the satellites which orbit interior to Miranda) is not currently understood.
However, it may be possible that after accreting from a planetary ring, Miranda at some point crossed the $3:1$~MMR with Umbriel \citep{Moons94}.

\cite{Tiscareno2013} determined that the location of transition from ring material to satellites at Uranus implies a critical Roche density of $1.2 \text{ g cm}^{-3}$.
Satellites with a bulk density of $\sim 1.2 \text{ g cm}^{-3}$ accreting from a Roche-interior ring in orbit around Uranus fall within the Torque-Dependent regime (see Figure~\ref{fig:boomerang_ss}).
Examination of Equation~\ref{eqn:dadt} shows that the evolution of a satellite depends upon the tidal love number $k_{2}$, and the tidal dissipation factor $Q$.
However these values are not well constrained for the Uraninan satellites.
For Uranus, $k_{2}$ is thought to be $\sim 0.104$ \citep{Murray:1999th}, while $Q$ could be as small as $500$ or as large as $10,000$ \citep{Lainey2016}.

\subsubsection{Formation of Miranda from an Ancient Massive Uranian Ring}
We hypothesize that Miranda accreted from an ancient ring orbiting Uranus $4$~Gy ago that was sufficiently massive for Lindblad torques to migrate the satellite from the FRL to an orbit beyond $a_{synch}$.
We further hypothesize that after migrating to an orbit beyond $a_{synch}$, tidal torques caused the satellite to migrate to its current orbit.
With this hypothesis in mind, we use the current semi-major axis of Miranda as a constraint in calculating the tidal ratio $k_{2}/Q$.
If we assume Miranda evolved to its current orbit from $a_{synch}$ over $4$~Gy from tidal torques alone, we may place a lower bound on the value of $k_{2}/Q$.
With these constraints we may integrate Equation~\ref{eqn:dadt} to determine $k_{2}/Q$.

Ignoring Lindblad torques, integration of Equation~\ref{eqn:dadt} for a satellite in a circular orbit yields:
\begin{equation}\label{eqn:a_F}
	a_{F}^{13/2} - a_{I}^{13/2} = \frac{39 M_{s} R_{p}^{5}}{2}\sqrt{\frac{G}{M_{p}}}\left(\frac{k_{2}}{Q}\right) \Delta t.
\end{equation}
Here $\Delta t$ is the amount of time tidal torques will cause a satellite to migrate from an initial semi-major axis $a_{I}$ to a final semi-major axis $a_{F}$.
If we assume that tidal torques caused Miranda to migrate from $a_{synch}$ to its current orbit in $4$~Gy, we may determine the required tidal ratio $k_{2}/Q$ for this to occur by rearranging Equation~\ref{eqn:a_F} to yield:
\begin{equation}\label{eqn:k2_Q}
	\frac{k_{2}}{Q} = \frac{2}{39M_{s}R_{p}^{5}\Delta t}\sqrt{\frac{M_{p}}{G}}\left[a_{F}^{13/2} - a_{I}^{13/2}\right].
\end{equation}
We set $a_{F}$ equal to Miranda's current semi-major axis, $a_{I} = a_{synch} \sim 3.26 R_{p}$, and $\Delta t = 4 \times 10^{9}$ years.
By substituting the mass and radius of Uranus into Equation~\ref{eqn:k2_Q}, as well as the current mass of Miranda, we find $k_{2}/Q \approx 3.3\times 10^{-5}$.
If $k_{2} = 0.104$ \citep{Murray:1999th}, this corresponds to a tidal dissipation factor of $\sim 3150$, and is well within the expected range for $Q$ \citep{Lainey2016}.

By definition, a satellite that accretes within the ``Torque-Dependent'' regime would be in a 2:1 resonance ($\mathcal{M} = 2$) with a location interior to the FRL when the satellite is located at the synchronous orbit.
Examining Equation~\ref{eqn:threshold}, we calculate the necessary surface-mass density of the ring for Lindblad torques to overcome the tidal torques.
We find that a Uranian ring with a surface-mass density of $\geq 17 \text{ g cm}^{-2}$ in a 2:1 resonance with a satellite located at $a_{synch}$ would be able to perturb the satellite to beyond $a_{synch}$.
At this point tidal torques would continue to migrate the satellite away from the planet.
This surface-mass density is on the order of the estimated surface-mass density of the Uranian rings today \citep{Esposito2006}.

\subsubsection{Using RING-MOONS to Model a Massive Uranian Ring}
In order to simulate the Uranian system in ``RING-MOONS'' we need to define the initial conditions of the system \citep{Hesselbrock2017}.
The physical characteristics of the particles in the ring and the accreted satellites, as well as the surface mass density profile of the ring all affect the outcome of the system.
As we are testing a hypothesis on the formation of Miranda, we assume the bulk density of the ring material, and any accreting satellites, to be identical to the bulk density of Miranda, $\rho_{s} = 1.2 \text{ g cm}^{-3}$ \citep{Tiscareno2013}.
Furthermore, we set the initial surface-mass density of the ring to follow a power law, such that $\sigma(r) = \sigma_{0}r^{-3}$, where $\sigma_{0}$ is a constant determined by the initial mass of the ring.
The ring extends from the upper atmosphere of Uranus to the FRL.
Lastly, we assume $k_{2} = 0.104$ \citep{Murray:1999th} and set $Q = 3000$.
This corresponds to a tidal ratio of $k_{2}/Q = 3.5 \times 10^{-5}$, which is slightly more than the calculated lower bound of $k_{2}/Q = 3.3\times 10^{-5}$.

\begin{figure}[t]
	\begin{center}
		\includegraphics[width=6.5in]{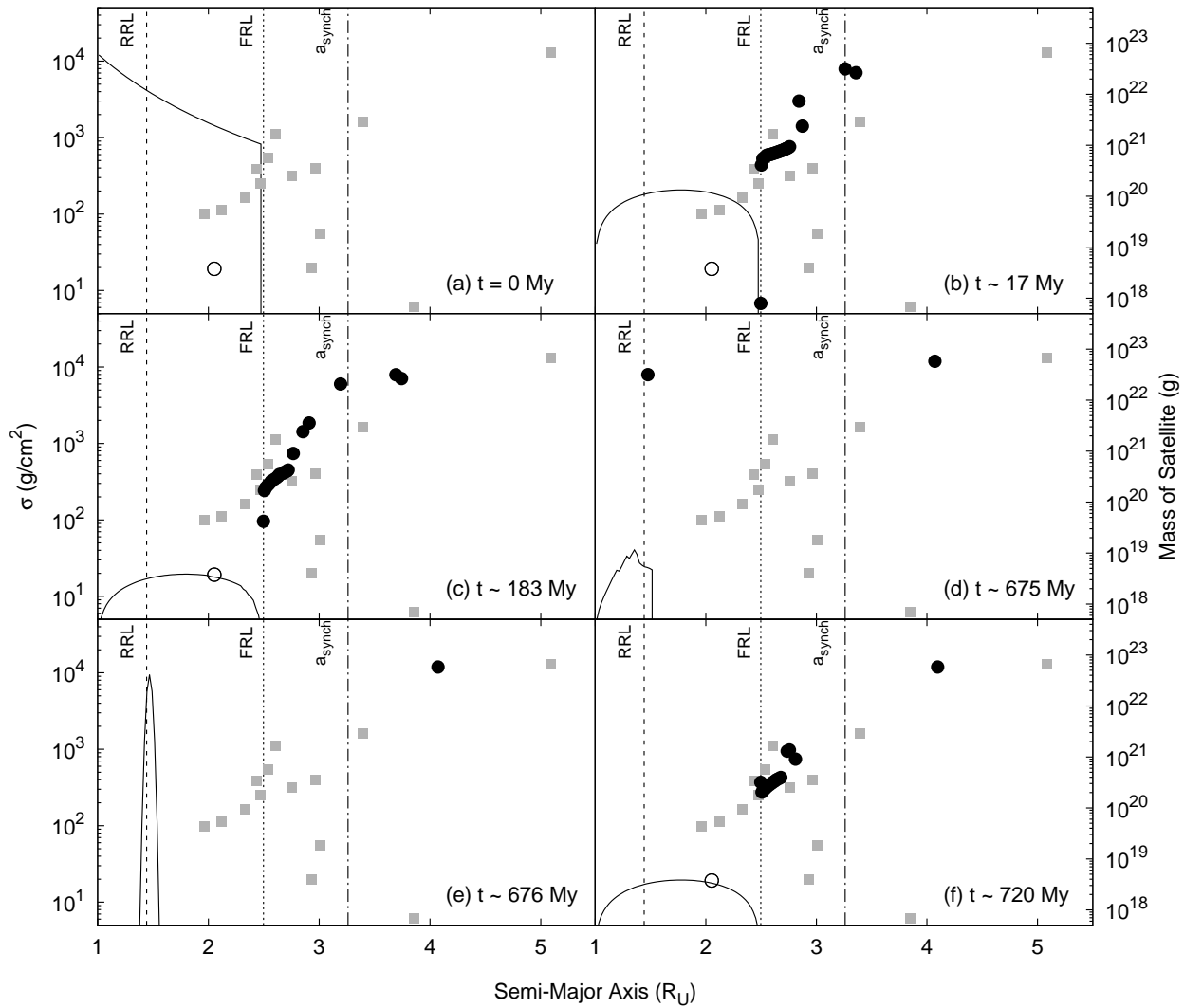}
		\caption{Evolution of the surface mass density of a Uranian ring with an initial mass of $3.0 \times 10^{23}$~g. The horizontal axis marks the distance from Uranus, the left vertical axis marks the surface-mass density of the ring (black line), and the right vertical axis marks the mass of the satellites. Solid black circles represent RING-MOONS satellites while the current satellite population is shown as gray squares. The locations of the RRL, FRL, and the synchronous orbit are marked with vertical lines. The open circle marks the required surface-mass density at the $\mathcal{M} = 2$ mode for Equation~\ref{eqn:threshold} to be true for a satellite at $a_{synch}$, ignoring all other modes. (a) Initial conditions. (b) Two satellites have evolved beyond $a_{synch}$ and the surface-mass density is above the threshold value. (c) The surface-mass density of the ring has fallen such that Equation~\ref{eqn:threshold} is no longer true. (d) The surface-mass density of the ring has declined.  The two satellites orbiting beyond $a_{synch}$ have merged into a Miranda-mass satellite.  The $18$ satellites interior to $a_{synch}$ have merged into one massive satellite that has migrated to the RRL. (e) The massive satellite has been disrupted at the RRL, generating a new ring which has begun to viscously spread. (f) The ring generated by the destruction of the massive satellite has accreted a new generation of $14$ satellites. However, the surface-mass density of the ring has fallen below the threshold value.  Overtime these satellites may gravitationally scatter to produce a system similar to the one observed today. }
		\label{fig:simulation}
	\end{center}
\end{figure}

In Figure~\ref{fig:simulation} we display the results from our RING-MOONS simulation.
The initial surface-mass density of a ring with a total mass of $3.0 \times 10^{23}$~g, as well as the current mass and location of the inner Uranian satellites is shown in Figure~\ref{fig:simulation}a.
Additionally, we have marked the semi-major axes of the RRL, FRL, and $a_{synch}$.
For a satellite to migrate beyond $a_{synch}$, the locations in the ring that are in resonance with the satellite must have a sufficient surface mass density to satisfy Equation~\ref{eqn:threshold}.
A satellite orbiting Uranus at $a_{synch}$ could be in resonance with both the $\mathcal{M} = 2$ and $\mathcal{M} = 3$ modes in the ring.
This permits multiple surface-mass density profiles for Equation~\ref{eqn:threshold} to remain true.
However, in Figure~\ref{fig:simulation}a we have marked the surface-mass density for the $\mathcal{M} = 2$ resonance mode to alone satisfy Equation~\ref{eqn:threshold} for a satellite at $a_{synch}$.

We find that the ring quickly transports ring material beyond the FRL where it is able to accrete into satellites \citep{Crida2012,Hesselbrock2017}.
Additionally, the surface mass density of the ring is initially sufficient for Lindblad torques to overcome the tidal torques, migrating the satellites away from the ring edge.
As displayed in Figure~\ref{fig:simulation}b, we see that after $17$~My the surface-mass density of the ring has decreased, yet remains sufficient to satisfy Equation~\ref{eqn:threshold}.
In $17$~My the ring has produced a collection of $25$ satellites, including two satellites which have migrated beyond $a_{synch}$.
Furthermore, Figure~\ref{fig:simulation}b shows that the surface mass density of the ring at the $\mathcal{M} = 2$ resonance location is sufficient for Lindblad torques to migrate the Miranda-mass satellite beyond $a_{synch}$.

The results of the simulation after $183$~My are displayed in Figure~\ref{fig:simulation}c.
At this point we see that the two exterior satellites have migrated beyond $a_{synch}$.
Once beyond the synchronous orbit, Equation~\ref{eqn:dadt} is positive as the tidal torques cause the two satellites to migrate away from the primary.
After $\sim 183$~My the ring has generated a total of $20$ satellites.
However, at this point the surface-mass density of the ring at the $\mathcal{M} = 2$ mode has fallen such that Equation~\ref{eqn:threshold} is no longer valid for a satellite located at $a_{synch}$.
We see that the surface-mass density of the ring has fallen such that Lindblad torques are unable to cause any other satellites to migrate beyond $a_{synch}$.
As the ring loses mass overtime, tidal torques cause the remaining $18$ satellites orbiting inside $a_{synch}$ to migrate inwards.

At $\sim 215$~My the two satellites orbiting beyone $a_{synch}$ merge to form a single Miranda-mass satellite.
The Miranda mass satellite continues to be migrated away from the primary, and reaches Miranda's current semi-major axis at $\sim 4.2 \times 10^{9}$ years.
As the simulation continues, the most massive satellite interior to $a_{synch}$ migrates inwards.
The tidal torque exerted on the massive satellite causes its semi-major axis to decrease more rapidly than any other satellites interior to $a_{synch}$.
Thus, as the massive satellite migrates inwards it accretes all the satellites interior to $a_{synch}$.
After $\sim 675$~My all satellites interior to $a_{synch}$ have merged into one massive body that has migrated to the RRL (see Figure~\ref{fig:simulation}d).

Upon reaching the RRL, the massive satellite has rapidly disrupted, with its material generating a new ring of material, in agreement with \citet{Leinhardt2012Disrupt}.
The ring begins to viscously spread as displayed in Figure~\ref{fig:simulation}e.
This ring deposits much of its mass onto Uranus, however it does spread material towards the FRL.
After $\sim 720$~My, the ring has spread material beyond the FRL forming a new generation of $14$ satellites interior to $a_{synch}$, which we show in Figure~\ref{fig:simulation}f.
The most massive satellite generated by the new ring is roughly half the mass of the satellite Puck.
However, as the mass of the ring decreases over time, the surface mass density of the new ring is insufficient to evolve any of these satellites beyond $a_{synch}$.

It is important to note that the value of $k_{2}/Q$ has a strong effect on the results of the simulation.
In Equation~\ref{eqn:threshold}, $Q$ and $\sigma(r)$ are inversely related.
If $Q \sim 10,000$, in contrast to the value we derived, the magnitude of the tidal torques is reduced.
In such a scenario, the surface-mass density of the ring for the Lindblad torques to overcome the tidal torques would be substantially less.
Thus, for the same mass ring, satellites would be more easily evolved to orbits beyond $a_{synch}$.
However, once the satellite has been migrated beyond $a_{Lind}$ tidal torques alone drive the migration of the satellite.
In this scenario, a Miranda-mass satellite would be unlikely to reach its current orbit within $4$~Gy due to the reduced tidal torque.
Conversely, if $Q \sim 500$, the necessary surface-mass density for Equation~\ref{eqn:threshold} to be true would be substantially greater.
Although such a ring would be able to evolve a satellite to the orbit of Miranda, the greater mass of the ring would result in a satellite much more massive than Miranda.

\subsection{Implications of Origin and Evolution of Miranda and Early Uranus}\label{sec:uranus_conclusion}

While the results displayed in Figure~\ref{fig:simulation} show that it is possible to evolve a Miranda-mass satellite to its current orbit from a Roche-interior ring, these results do not fully reproduce the system as we see it today.
There are 13 Uranian satellites interior to $a_{Lind}$, and our RING-MOONS results do not reproduce all of the inner Uranian satellites.
The most difficult satellites to model with RING-MOONS are the satellites Mab and Puck.
We find that rings that are massive enough to perturb two bodies beyond $a_{synch}$ typically produce satellites much more massive than Miranda and/or Puck.

There are several dynamical processes that are not modeled in RING-MOONS, but which may be important for reproducing the architecture of the inner Uranian satellite system.
RING-MOONS treats all satellite-satellite interactions as direct mergers, whereas in reality close encounters could also cause satellites to disrupt, or scatter \citep{Hesselbrock2017, Leinhardt:2012kd, Salmon2017}.
This causes RING-MOONS to produce systems where satellite mass generally increases with semi-major axis, and prevents the model from producing small satellites that could be scattered into distant orbits.
This may explain why the architecture of the inner satellite system in Figure \ref{fig:simulation}f does not match the observed system, especially for the satellites Perdita and Cupid (Figure \ref{fig:architecture}).
Furthermore, satellites accreting from a Roche-interior ring are often in near resonance with each other.
A satellite orbiting beyond $a_{Lind}$ will no longer directly exchange angular momentum with the ring.
However, if an interior satellite that is still in resonance with the ring enters into resonance with the exterior satellite, the Lindblad torque exerted onto the interior satellite will be passed on to the exterior satellite.
Thus, Lindblad torques may evolve a satellite to an orbit beyond $a_{Lind}$ through a Laplace resonance chain \citep{Salmon2017}.

The limitations to satellite-satellite dynamics as modeled in RING-MOONS presents an additional difficulty when considering the major Uranian satellites, Ariel, Umbriel, Titania, and Oberon.
The orbital migration of these bodies can have strong effects on each other, and also potentially an inner satellite system as well.
In this work, we assume the major Uranian satellites formed soon after Uranus itself and are primordial.
Thus, the orbital migration of the inner satellites which form in the Torque-Dependent regime differs from the orbital migration of the major satellites.
However, the gravitational influence the major satellites on the architecture of the inner satellite system as displayed in Figure \ref{fig:simulation} needs to be investigated further.
We find that Miranda, in its outward migration, would encounter several mean-motion resonances with Ariel and Umbriel.
Indeed these interactions may have left Miranda with the inclination observed today \citep{Tittemore90, Moons94}.

Reproducing the Uranian system as it is observed today is a significant challenge due to a number of factors.
The inner Uranian satellites are currently tightly packed, leading to a highly dynamic, chaotic system \citep{French2012}.
Many of these satellites experience a combination of mean-motion eccentricity and inclination resonances, making long term orbital integrations of the system difficult \citep{French2015}.
Furthermore, the satellites themselves are expected to have experienced multiple disruptive impacts with heliocentric material.
Many of these collisions would likely be catastrophic to the inner satellites.
It is thought that the system observed today has been collisionally evolved on a timescale of $10^{8}$ years \citep{Colwell92}.
\citet{Colwell92} argue that the ring-satellite system we observe today are leftover collision fragments from some older population of satellites. 
This makes it extremely challenging to even hypothesize which satellites existed millions to billions of years ago.

Lastly, the assumption that any satellite that has been migrated to the RRL would be tidally disrupted may be an over simplification.
The RRL marks the location where a cohesionless, strengthless object would be tidally disrupted, and is dependent upon the satellite bulk density.
As discussed in Section \ref{sec:Uranus}, the bulk density of the inner Uranian satellites is indirectly constrained from the rings \citep{Tiscareno2013}, but our estimate to the location of the RRL may not be correct for every inner satellite.
Furthermore, the inner Uranian satellites likely have some form of internal cohesion, which would prevent them from being tidally disrupted at the RRL \citep{Black2015}.
The uncertainty in the location of tidal breakup does not affect our result for the formation of Miranda, but it does affect our results for satellites which accrete from rings created by tidally disrupted satellites (Figure \ref{fig:simulation}f).
If the location of tidal breakup is inward of the RRL, the subsequently formed ring would transport a smaller mass of material beyond the FRL to form satellites \citep{Hesselbrock2017}.
Thus, the mass of the satellite system shown in Figure \ref{fig:simulation}f serves as an upper bound on the mass of the inner satellite system.

Due to these complications, connecting the results of RING-MOONS depicted in Figure~\ref{fig:simulation} to the system observed today is not straightforward.
The population of satellites interior to Miranda as produced by RING-MOONS and depicted in Figure~\ref{fig:simulation}f has a total mass of $\sim 6.6 \times 10^{21}$~g.
This is agrees with the current population of satellites interior to Miranda, which are estimated to have a total mass of $\sim 6.5 \times 10^{21}$~g.
We expect that heliocentric impacts with the inner satellites generated by RING-MOONS would collisionally evolve the system to produce the ring-satellite system of Uranus observed today \citep{Colwell92,Colwell93}.
Nevertheless, our results do give some constrains on the origin of Miranda, and are in broad agreement with the total mass of satellites interior to Miranda, even if we do not reproduce the details of their orbital architecture.

\section{Conclusion}\label{sec:conclusion}

In this work we have developed a model to investigate the evolution of coupled ring-satellite systems.
Satellites in orbits near massive planetary rings may exchange angular momentum with both the ring and the primary body.
The dynamics of how angular momentum is exchanged with the satellite creates three distinct evolution regimes for satellites accreting from massive planetary rings.
We term these the Boomerang, Slingshot, and Torque-Dependent regimes.
The three formation regimes, outlined in Section~\ref{sec:regimes}, provide a method to analyze the dominant dynamics of massive ring systems interacting with nearby satellites.
Identifying in which regime a ring-satellite system exists enables us to hypothesize the past and future evolution of the system.
Furthermore, our model makes predictions of the behavior of many systems and motivates future searches for potential rings and/or satellites.

In the Slingshot regime, the synchronous orbit lies inside the location of the FRL.
If a ring in the Slingshot regime is able to viscously spread material to the FRL, the system should produce at least one satellite that would likely be observable today as both the ring and tidal torques would cause the satellite to migrate away from the primary.
In the Boomerang regime the synchronous orbit lies outside the maximum orbit Lindblad torques could perturb a satellite.
Satellites that form out of massive rings in the Boomerang regime may migrate away from the primary via Lindblad torques, however over time tidal torques cause the satellite to migrate inwards.
As a satellite's semi-major axis decreases, tidal stresses across the body increase and may disrupt the satellite into forming a new ring.
As shown in HM17, it is possible for a planetary body in the Boomerang regime to have a cycle of ring formation and satellite accretion that persists for billions of years.
The dynamics of the coupled ring-satellite systems in the Boomerang regime provide a mechanism to repeatedly generate satellites, and rings, provided satellites disrupt interior to the FRL \citep{Black2015,Hesselbrock2017}.
An observation of a system that is in the Boomerang regime with a satellite inside $a_{synch}$ is an indication that a ring may have existed, or currently exists, at the system.
Satellites in Boomerang regime systems (such as at Eris, see Figure~\ref{fig:boomerang_ss}) could motivate a search for rings.

In the Torque-Dependent regime the synchronous orbit lies between the FRL and the maximum orbit Lindblad torques could perturb a satellite.
Planetary systems with rings in the Torque-Dependent regime may exhibit characteristics of both Boomerang and Slingshot systems.
Much like satellites that accrete from a ring in the Boomerang regime, satellites in the Torque-Dependent regime experience a competition between Lindblad and tidal torques.
As Torque-Dependent systems evolve over time, they may transition from having rings that are sufficiently massive to cause satellites to migrate beyond the synchronous orbit (Slingshot), to systems in which tidal interactions dominate the migration of ring-accreted satellites (Boomerang).
Rings in the Torque-Dependent regime may initially be massive enough to cause a satellite that accretes at the FRL to migrate beyond the synchronous orbit, however over time the mass of the ring may deplete such that tidal torques dominate a satellite's migration.
Using RING-MOONS, we have shown that the Uranian ring-satellite system may have experienced such an evolution.
Our results show that Miranda may have accreted from an ancient ring in orbit around Uranus that was sufficiently massive to cause the satellite to migrate to its current orbit.
Torque-Dependent regime systems like Uranus may repeatedly generate satellites and rings, similar to systems in the Boomerang regime.
As the mass of the Uranian ring depleted, any satellites interior to Miranda could not reach the synchronous orbit, and eventually migrated back toward Uranus, possibly going through complex phases of scattering, disruption, and reaccretion.
Thus, satellites identified in Torque-Dependent regime systems (such as at Quaoar, see Figure~\ref{fig:boomerang_ss}) would also motivate a search for potential rings or interior satellites.

\clearpage
\section{Figures}\label{sec:figures}
\newpage

\newpage

\newpage

\newpage

\clearpage
\software{RING-MOONS \citep{Hesselbrock2017}}



\end{document}